\documentclass[twocolumn,preprintnumbers,amsmath,amssymb,prl]{revtex4}

\usepackage{graphicx}

\begin{document}

\title{Experimental demonstration of quantum teleportation of broadband squeezing}

\author{Hidehiro Yonezawa$^{1,2}$, Samuel L.\ Braunstein$^{3}$
and Akira Furusawa$^{1,2}$}
\affiliation{$^{1}$Department of Applied Physics, School of Engineering,
The University of Tokyo,\\
7-3-1 Hongo, Bunkyo-ku, Tokyo 113-8656, Japan \\
$^{2}$CREST, Japan Science and Technology (JST) Agency,
1-9-9 Yaesu, Chuo-ku, Tokyo 103-0028, Japan \\
$^{3}$Computer Science, University of York, York YO10 5DD, UK}

\begin{abstract}
We demonstrate an unconditional high-fidelity teleporter capable of
preserving the broadband entanglement in an optical squeezed state.
In particular, we teleport a squeezed state of light and observe
$-0.8 \pm 0.2$dB of squeezing in the teleported (output) state. We show that
the squeezing criterion translates directly into a sufficient criterion 
for entanglement of the upper and lower sidebands of the optical field.
Thus, this result demonstrates the first unconditional teleportation 
of broadband entanglement. Our teleporter achieves sufficiently high
fidelity to allow the teleportation to be cascaded, enabling, in
principle, the construction of deterministic non-Gaussian operations.
\end{abstract}

\maketitle

One of the most significant developments in quantum optics in recent years,
which promises to revolutionize the field, has been the move from on-line
evolution to the use of off-line resources with detection and feed-forward
\cite{Mertz90} to achieve the same (and often superior) result.
Here, on-line refers to the direct unitary evolution of a signal, as
contrasted with the use of an auxiliary quantum system (an off-line 
resource) that is coupled to the orginal signal and whose measurement 
yields some classical information about that signal. This information 
is processed off-line and used to modify the signal (in a feed-forward 
step).

This approach was demonstrated, for example, in the making of a near-ideal
phase-insensitive amplifier \cite{Josse06}, an achievement that would
be virtually impossible with the direct (on-line) use of a laser
amplifier. Similarly, squeezing of anything other than vacuum states
has been virtually impossible until the recent realization that the
squeezed resources could be moved off-line and replaced with a
feed-forward scheme \cite{Yoshikawa07,Filip05}. This opens the door 
for near-ideal quantum non-demolition measurements that have long been 
championed for gravity-wave detection and quantum metrology more generally
\cite{Filip05}. In fact, one of the earliest examples of a feed-forward
protocol is quantum teleportation \cite{Bennett93,Vaidman94}. Here there is an
entangled (off-line) resource, Bell-state detection and feed-forward.
In ideal teleportation the teleported state reproduces the input state,
thus implementing the identity evolution or gate. It was later realized
that by modifying the entangled resource, other quantum gates could
be implemented \cite{Gottesman99}.  While this applies equally to both
discrete and continuous variable (CV) systems, the latter's ability to
construct non-Gaussian gates would supplement linear optics and deliver
universal quantum computation in the CV setting. In particular, a
cascaded teleportation protocol would allow for the construction of
a (non-linear) cubic phase gate \cite{Gottesman01,Furusawa07}.

Since the first demonstration of teleportation in the late 1990s 
\cite{topten}, we have seen remarkable progress in teleportation related 
technologies. In particular, in the CV setting, experiments teleporting 
coherent states of light 
\cite{Furusawa98,Bowen03a,Zhang03,Takei05e,Yonezawa04} and matter
\cite{Sherson06}, as well as halves of entangled states (entanglement 
swapping) \cite{Takei05e} and squeezed states \cite{Takei05s} have 
all been reported.  In the entanglement swapping experiment the 
entanglement was shared between two spatially separated modes. By contrast, 
in the experiment reported here the entanglement lies between upper and 
lower sidebands of the same spatial mode 
\cite{JZhang03,Huntington02}.  Hence the entire entangled 
state is teleported. In the experiment teleporting squeezed states 
\cite{Takei05s}, the states were teleported with a fidelity exceeding the 
classical limit \cite{Braunstein00}. Nonetheless, squeezing itself was 
not preserved in the output state \cite{Takei05s}.  By contrast, the 
teleportation reported here maintains the squeezing and the entanglement 
in the output state. This is a prerequisite for any cascaded teleportation
protocol, and indeed for any multi-step information processing scheme.

For the CV teleportation experiments described above (including entanglement
swapping), no more than $-3$dB of squeezing has been required. However,
in order to maintain squeezing at the output, at least $-4.8$dB of 
squeezing is essential. While high levels of squeezing have been generated
in the past (recently up to $-9$dB \cite{Takeno07}), we report here the
first {\it application} of such high levels of squeezing. This represents
a significant breakthrough in phase stabilization.

It is not surprising that the teleportation of an entire entangled state, 
as reported here, necessarily entails the preservation of squeezing in the 
(output) teleported state.  In fact, when the squeezing criterion is 
decomposed into quadrature amplitudes of the upper and lower 
sidebands, it translates directly into a sufficient criterion for
entanglement between these sidebands \cite{Duan00,footnoteBB}
\begin{eqnarray}
          \Delta_{\rm sq} &\equiv&
          \langle (
                   \Delta [ \hat x(\Omega_s) +\hat x(-\Omega_s) ] )^2
                                \rangle  \nonumber \\
&&+ \langle ( \Delta [ \hat p(\Omega_s) -\hat p(-\Omega_s) ] )^2 \rangle
< 1\;.
\label{BBcriterion}
\end{eqnarray}
Here $\Delta \hat O$ denotes the uncertainty in $\hat O$, 
$\Omega_s$ is the sideband frequency, and
$\hat x(\pm \Omega_s)$ and $\hat p(\pm \Omega_s)$ are canonically
conjugate quadrature operators for the two sideband modes, obeying the
uncertainty relation 
$\langle [ \Delta \hat x (\pm \Omega_s) ]^2 \rangle
\langle [ \Delta \hat p (\pm \Omega_s) ]^2 \rangle \geq \frac{1}{16}$
(with $\hbar=\frac{1}{2}$) \cite{JZhang03,footnoteBB}. 
In our experiment, we estimate a value of $\Delta_{\rm sq}=0.83 \pm 0.04$
for the teleported states ($\Delta_{\rm sq}=0.24 \pm 0.01$ for the input
state), demonstrating successful unconditional quantum teleportation 
of entanglement. This is the first such {\it unconditional} demonstration
for either discrete or continuous quantum variables.

The CV quantum optical field may be described in terms of the annihilation
operator for the electromagnetic field $\hat a(t)$ or its Fourier transform
$\hat a(\Omega)$, here written in the rotating frame about the central
optical frequency. From these the sideband quadrature operators may be
defined in terms of the annihilation and creation operators for the
sideband modes as
$\hat x (\Omega)=\frac{1}{2}[ \hat a(\Omega) +\hat a^\dagger (\Omega)]$ and
$\hat p (\Omega)= \frac{i}{2}[ \hat a^\dagger (\Omega)-\hat a(\Omega)]$.
These sideband quadratures satisfy the usual canonical commutation relations
$[\hat{x}(\Omega), \hat{p}(\Omega')]=\frac{i}{2}\delta(\Omega-\Omega')$.
In the experiment, we treat the four sideband operators
$\hat x(\pm \Omega_s)$ and $\hat p(\pm \Omega_s)$ at the sideband frequency
$\Omega_s$ or combinations of these operators. Homodyne detectors measure
$\hat x_{\rm hom}(t) \equiv \hat a (t) +\hat a^\dagger (t)$ which in
the frequency (or RF) domain has the form \cite{JZhang03}
\begin{eqnarray}
\hat x_{\rm hom}(\Omega) &\equiv & \hat a (\Omega) +\hat a^\dagger (-\Omega)
\nonumber \\
&=& \hat x(\Omega)+\hat x(-\Omega) +i [ \hat p(\Omega)-\hat p(-\Omega) ] \;.
\end{eqnarray}
The RF in-phase (real) and RF out-of-phase (imaginary) parts of 
$\hat x_{\rm hom}(\Omega)$ may be extracted simultaneously and are used
for the feed-forward step of the teleportation. In addition, a spectrum
analyzer is used to measure the noise power (the sum of the variances of
both real and imaginary components) of this homodyne 
signal \cite{SpectrumAnalyser}
\begin{equation}
\langle ( \Delta [\hat x(\Omega) +\hat x(-\Omega) ]   )^2  \rangle
+ \langle ( \Delta [\hat p(\Omega) -\hat p(-\Omega) ] )^2  \rangle \;.
\end{equation}
At the sideband frequency $\Omega=\Omega_s$ this noise power is just
$\Delta_{\rm sq}$ from our entanglement criterion~(\ref{BBcriterion}).
Thus, when the noise level in the spectrum
analyzer shows broadband squeezing below the vacuum level, it also
indicates the presence of entanglement between upper and lower sidebands.
In other words, if the teleported state preserves squeezing below the
vacuum level, then the teleported state also preserves entanglement
between the sidebands.

       \begin{figure}[ht]
         \begin{center}
            \includegraphics[width=0.73\linewidth]{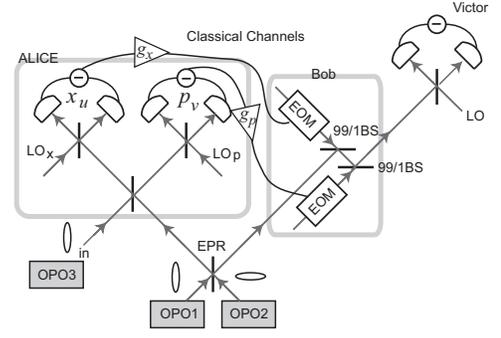}
\caption{
Experimental setup for CV quantum teleportation.
OPOs are optical parametric oscillators.
EOMs are electro-optical modulators.
All beam splitters except those labeled as 99/1 BSs are 
50/50. LOs are local oscillators for homodyne detection.}
            \label{fig-setup}
         \end{center}
\end{figure}

Figure~\ref{fig-setup} shows the experimental setup for CV quantum
teleportation of an electromagnetic field mode. 
We use three optical parametric oscillators (OPOs) which contain
periodically poled KTiOPO$_4$ as a nonlinear medium.
The output of a Ti:sapphire laser at 860nm is frequency doubled in
an external cavity containing a 10mm long potassium niobate crystal.
The output beam at 430nm is divided into three beams to pump three OPOs.
The pump powers are around 120mW for OPO3 and around 100mW for OPO1 and
OPO2. OPO1 and OPO2 are used to generate the EPR beams. The third OPO is
used to generate the quadrature-squeezed input state (except when
teleporting a coherent state, in which case the input is generated by
modulating a weak coherent beam from the laser at frequency sidebands of 1MHz).

Here we summarize the broadband description of teleportation as may 
be found in Ref.~\onlinecite{vanLoock00}, 
but other than the mixing of upper and
lower sidebands in the original shared EPR entangled states and in the
homodyne detection, it is very close to the single mode 
description \cite{Braunstein98,Furusawa98,Bowen03a,Zhang03,Takei05e,Takei05s}.
In particular, CV teleportation requires one of the EPR beams to be sent
to Alice ($\hat{x}_{A}(\Omega),\hat{p}_{A}(\Omega)$), and the other to Bob 
($\hat{x}_{B}(\Omega),\hat{p}_{B}(\Omega)$). Alice makes a joint homodyne
measurement between her EPR beam ($\hat{x}_{A}(\Omega),\hat{p}_{A}(\Omega)$)
and the input beam ($\hat{x}_{\rm in}(\Omega),\hat{p}_{\rm in}(\Omega)$),
and sends her homodyne measurement results ($x_u(\Omega), p_v(\Omega)$)
to Bob through classical channels. Bob receives Alice's measurement
results and ``displaces'' his EPR beam
($\hat{x}_{B}(\Omega),\hat{p}_{B}(\Omega)$) to reconstruct the input state.

The gains of the classical channels are defined as
$g_x=\langle\hat{x}_{\rm out}\rangle/\langle\hat{x}_{\rm in}\rangle$ and
$g_p =\langle\hat{p}_{\rm out}\rangle/\langle\hat{p}_{\rm in}\rangle$,
respectively (over the same RF bandwidths), where
($\hat{x}_{\rm out},\hat{p}_{\rm out}$) describes the output mode. We set
the gains to near unity, obtaining $g_x=1.00 \pm 0.01$ and
$g_p = 1.00 \pm 0.01$ (see Ref.~\onlinecite{Zhang03} 
for the tuning
procedure). These gains remained fixed throughout the experiment. The
visibilities of the homodyne detectors were about 98\%.

       \begin{figure}[th]
                    \includegraphics[width=\linewidth]{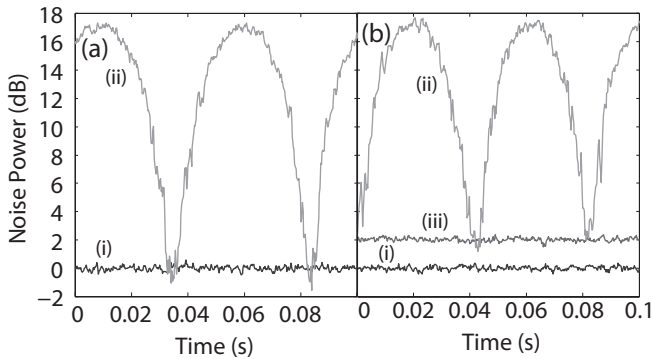}
\caption{
Quantum teleportation of a coherent state.
Plots (a) and (b) show the measurement results by a spectrum analyzer.
Plot (a) shows the input coherent state. Trace (i) shows vacuum noise level.
Trace (ii) shows the input coherent state with the phase scanned. Plot (b)
shows the output state of the teleportation for the $x$ quadrature ($p$
quadrature not shown). Trace (i) plots the vacuum noise level. Trace (ii)
plots the results of the teleportation with the phase of the input coherent
state scanned. Trace (iii) plots the variance for teleported vacuum. All
traces except traces (ii) are averaged 30 times. The center frequency is
1MHz.  The resolution and video bandwidths are 30kHz and 300Hz, respectively.
}
\label{fig2}
         \end{figure}

For unity gain and no losses, the output mode may be written 
as \cite{vanLoock00,Takei05e}
\begin{eqnarray}
\hat{x}_{\rm out}(\Omega_s)&=&\hat{x}_{\rm in}(\Omega_s)-
[\hat{x}_{A}(\Omega_s)-\hat{x}_{B}(-\Omega_s)] \nonumber \\
\hat{p}_{\rm out}(\Omega_s)&=&\hat{p}_{\rm in}(\Omega_s)+
[\hat{p}_{A}(\Omega_s)+\hat{p}_{B}(-\Omega_s)]\;.
\end{eqnarray}
In the ideal case, EPR beams satisfy
$\hat{x}_{A}(\Omega_s)-\hat{x}_{B}(-\Omega_s)\rightarrow 0$ and
$\hat{p}_{A}(\Omega_s)+\hat{p}_{B}(-\Omega_s) \rightarrow 0$, yielding
a teleported state that is identical to the input state. Treating all three
squeezed states as pure with the same squeezing parameter $r$, the
variances of the teleported state become
\begin{equation}
\langle ( \Delta \hat{x}_{\rm out} )^2 \rangle=\frac{3}{4} e^{-2r}\;,\quad\!\!
\langle ( \Delta \hat{p}_{\rm out} )^2 \rangle
=\frac{e^{2r} + 2\, e^{-2r}}{4} \;,
\end{equation}
compared with the variances of a vacuum mode \cite{footnoteBB}
$\langle ( \Delta \hat{x}_{\rm vac} )^2 \rangle =
\langle ( \Delta \hat{p}_{\rm vac} )^2 \rangle = \frac{1}{4}$.
Therefore in order to observe squeezing at the output, $e^{-2r}<\frac{1}{3}$
should be satisfied, corresponding to at least $-4.8$dB squeezing.

As a means of calibrating the performance of our teleporter, we use the mean 
fidelity $F_{\rm coh}=\langle \alpha |\rho_{\rm out}|\alpha\rangle$ for
teleporting coherent states (of coherent amplitude $\alpha$) yielding
output states $\rho_{\rm out}$. For a coherent state and unity gain, 
the fidelity can then be estimated from the variances of the 
outputs \cite{Takei05e}
\begin{equation}
F_{\rm coh}=\frac{2}{\sqrt{[1+4\langle(\Delta\hat{x}_{\rm out})^2\rangle]
[1+4\langle(\Delta\hat{p}_{\rm out})^2\rangle]}} \;.
\label{eq:Fcoh}
\end{equation}
Figure~\ref{fig2} shows the results of teleportation of a coherent state. 
Plots a and b show the noise power of the input and output states 
(for the $x$ quadrature; $p$ quadrature not shown), 
measured by a spectrum analyzer.  We use EPR beams
created from a pair of squeezed vacua of around $-6$dB each. 
Ideally, that level of squeezing would yield a fidelity $F_{\rm coh}=0.8$, 
and would enable cascading the teleportation scheme four
times \cite{Suzuki06}.  If cascaded, such a high-fidelity teleporter 
with a photon number detector would then allow for the construction of
a deterministic non-Gaussian gate like a cubic phase
gate \cite{Gottesman01,Furusawa07} for CVs.

In fact, achieving such a high fidelity teleporter poses significant
technical challenges. Indeed, although high levels of squeezing are in
themselves feasible (recently up to $-9$dB \cite{Takeno07}),
such high levels of squeezing in a real experiment 
(such as teleportation) would require very good phase locking.
As mentioned above, to realize our high-fidelity teleporter, we were 
able to generate and use $-6$dB squeezing. For experiments
using high squeezing levels, the anti-squeezed quadrature 
easily contaminates the squeezed quadrature \cite{Takeno07}. Hence,
unusually high mechanical stability of the experimental setup was needed 
to avoid such degradation in the squeezing.

To measure the input state, we block Alice's EPR beam and lock 
Alice's two homodyne detectors to the same quadrature simultaneously. 
The difference current from her two detectors was fed into a spectrum 
analyzer. In these measurements, the phase of the input coherent state was 
scanned. The maximum noise amplitude of the input state was found to be 
around $17$dB (corresponding to a coherent amplitude of $\alpha\simeq 3.5$). 
In Fig.~\ref{fig2}(b) the maximum noise amplitude of the output of the 
teleported state was also around $17$dB, confirming the near unit-gains of 
the classical channels. The variances of the output quadratures were 
measured separately as \cite{footnoteBB},
$\langle (\Delta \hat{x}_{\rm out} )^2\rangle = 2.0 \pm 0.2$dB
and $\langle (\Delta \hat{p}_{\rm out} )^2\rangle =2.3 \pm 0.2$dB
relative to the vacuum. These results are in good agreement with
theoretical values $2.1$dB and $2.2$dB which are calculated from
experimental losses and separately measured EPR correlations 
$\langle[\Delta(\hat{x}_{A}-\hat{x}_{B})]^2\rangle=-5.6\pm0.2$dB and 
$\langle[\Delta(\hat{p}_{A}+\hat{p}_{B})]^2\rangle=-5.5\pm0.2$dB.
Using the above variances and Eq.~(\ref{eq:Fcoh}), we obtain near 
ideal fidelity $F_{\rm coh}=0.76 \pm 0.02$ for our teleporter, 
exceeding the classical limit 
$F_{\rm coh}=\frac{1}{2}$ \cite{Braunstein00,Braunstein01,Hammerer05}, 
the no-cloning limit $F_{\rm coh}=\frac{2}{3}$ \cite{Grosshans01} 
and the highest fidelity reported to date \cite{Takei05e}.

       \begin{figure*}[ht]
      \begin{tabular}{cc}
                \begin{minipage}{0.46\hsize}
                 \begin{flushright}
                   \includegraphics[width=0.61\linewidth]{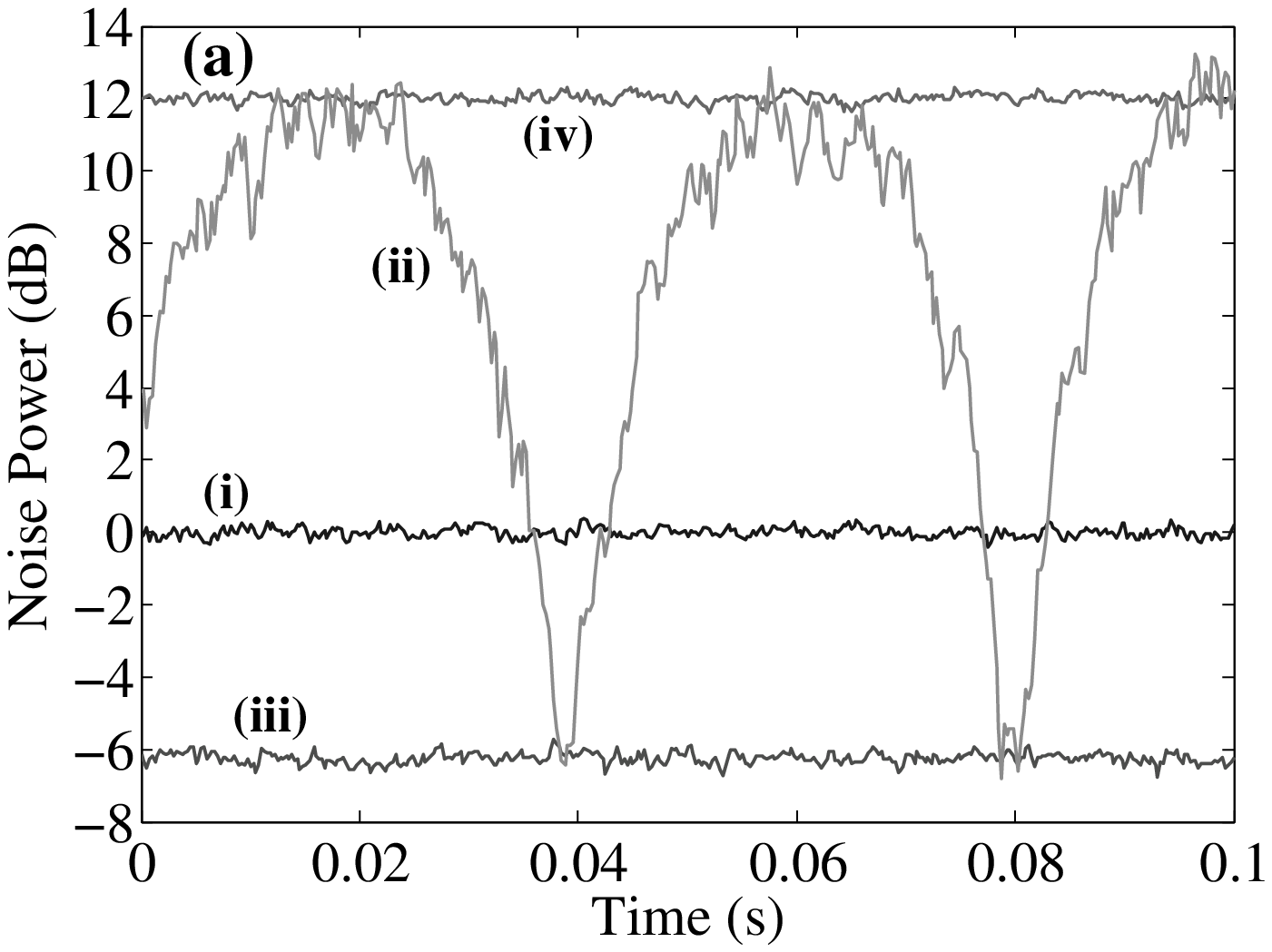}
                 \end{flushright}
                \end{minipage}
                \begin{minipage}{0.46\hsize}
                 \begin{flushleft}
                   \includegraphics[width=0.61\linewidth,clip]{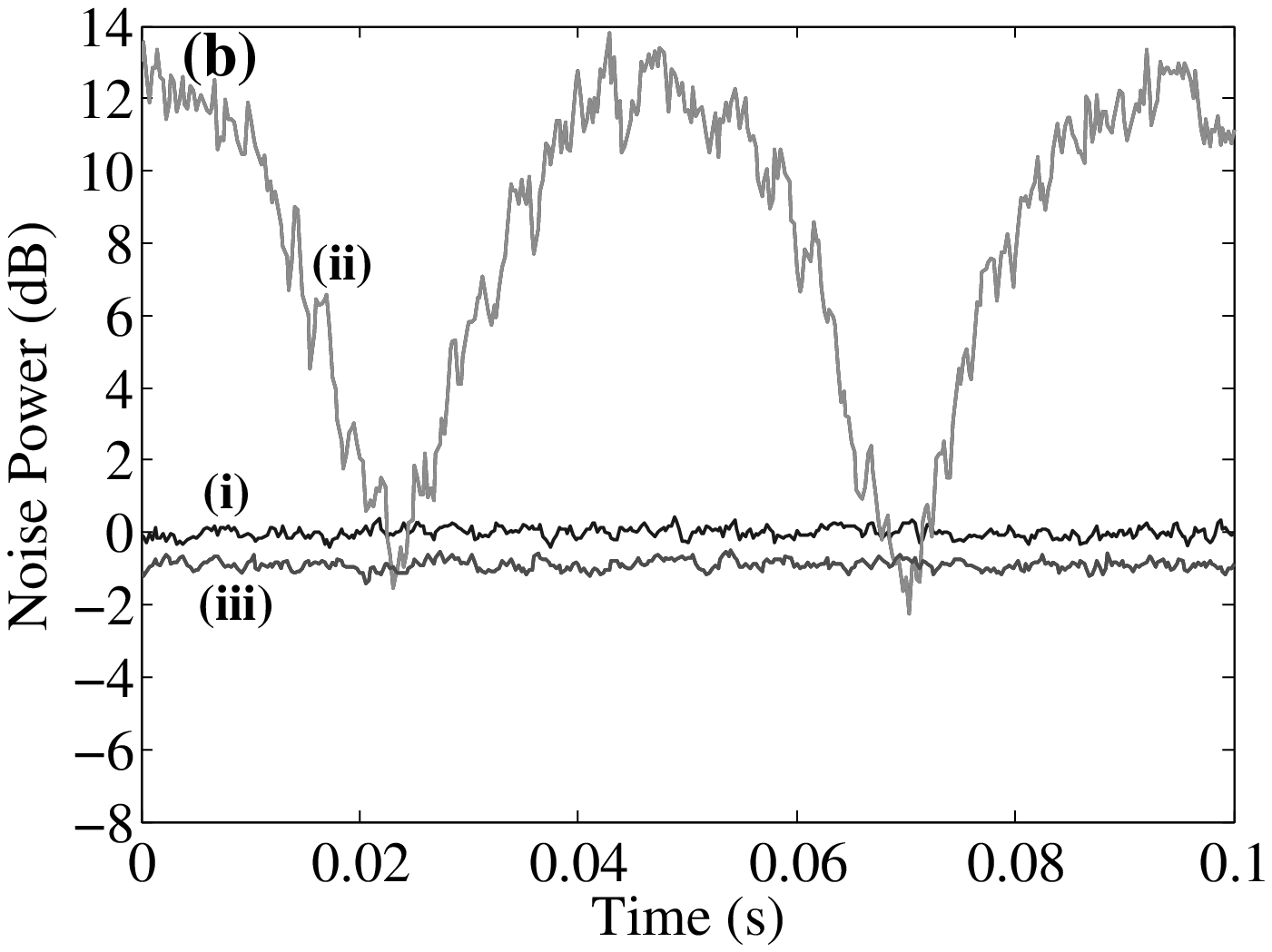}
                 \end{flushleft}
                \end{minipage} \\
                \begin{minipage}{0.46\hsize}
                 \begin{flushright}
                   \includegraphics[width=0.61\linewidth]{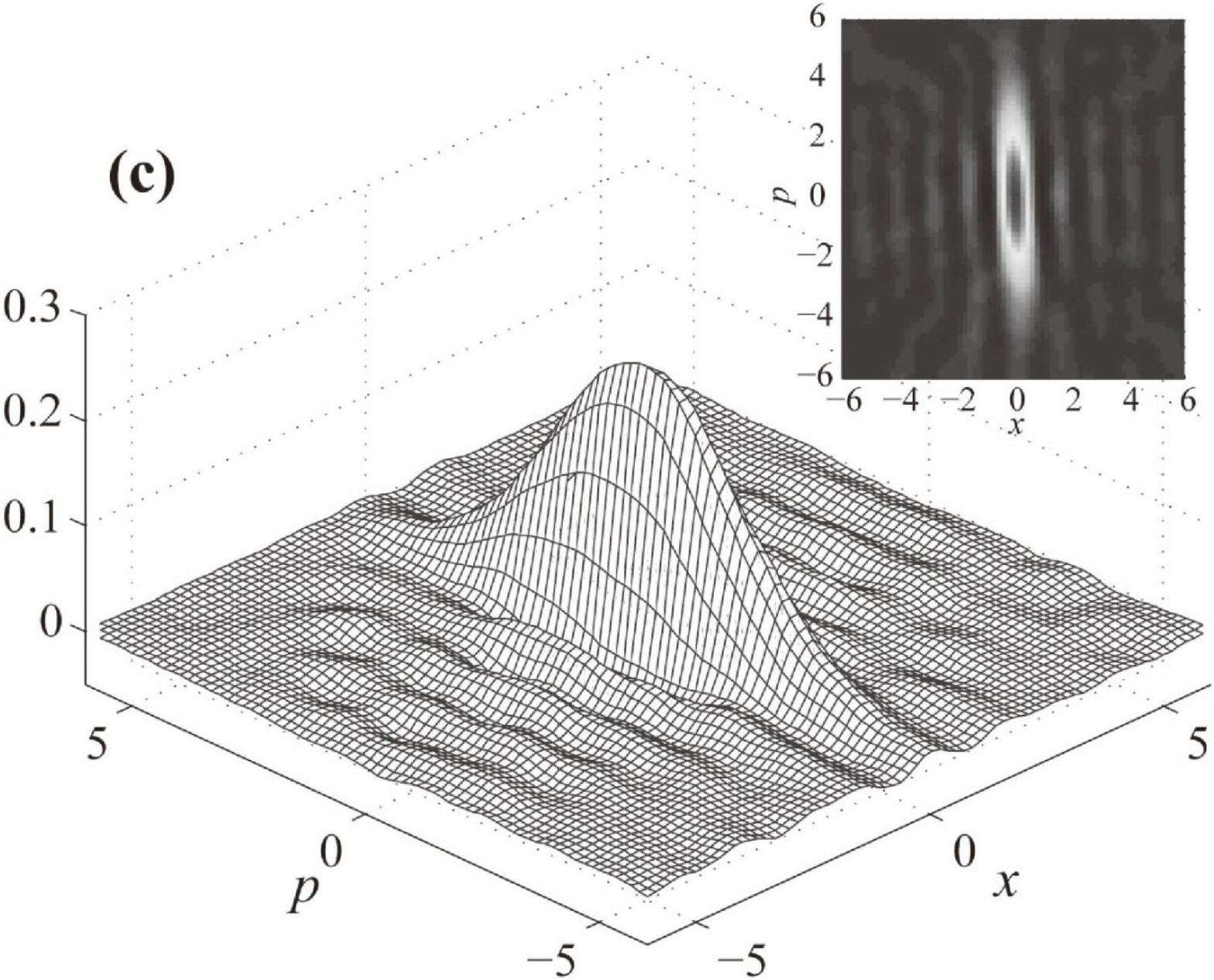}
                 \end{flushright}
                \end{minipage}
                \begin{minipage}{0.46\hsize}
                 \begin{flushleft}
                   \includegraphics[width=0.61\linewidth,clip]{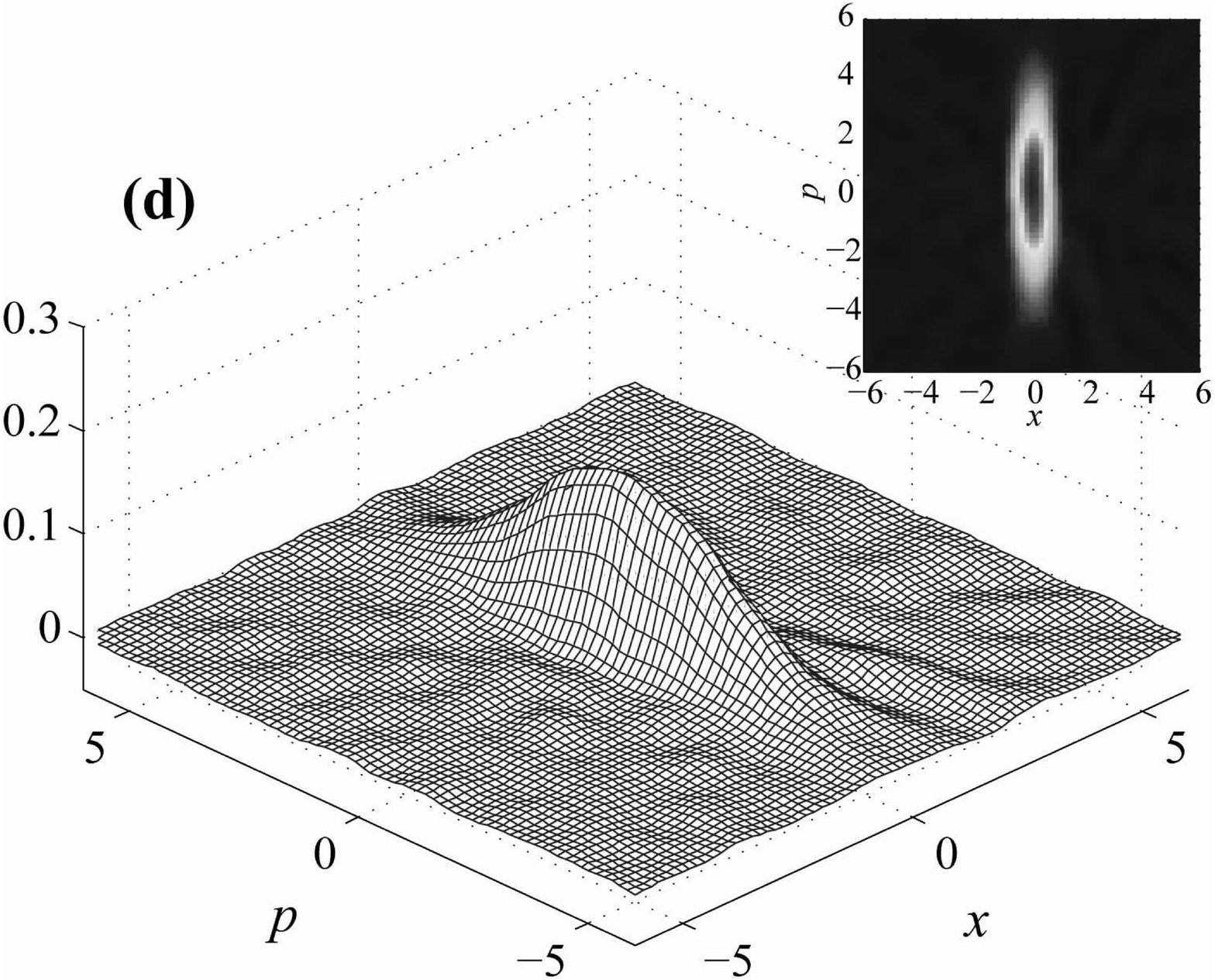}
                 \end{flushleft}
                \end{minipage}
              \end{tabular}
\caption{
Quantum teleportation of a squeezed state. 
Plots (a) and (b) show the measurements from a spectrum analyzer. 
Plot (a) shows the squeezed state to be teleported. Trace (i) plots
the vacuum noise level. Trace (ii) plots the squeezed state with 
phase scanned. Trace (iii) and (iv) show the squeezed state with 
phase locked to the squeezed and anti-squeezed quadratures. Plot (b)
shows the output state of the teleportation for the $x$ quadrature
($p$ quadrature not shown). Trace (i) plots the vacuum noise level.
Trace (ii) plots the teleported state with the input state's
phase scanned. Trace (iii) plots the teleported state
with the input state's phase locked to the $x$ quadrature. All traces
except traces (ii) are averaged 30 times. Plots (c) and (d)
show the Wigner functions of the input and output states of the
teleportation, respectively.}
\label{fig3}
       \end{figure*}

We now describe the teleportation of a squeezed state. Figure~\ref{fig3}(a)
shows the input squeezed state measured by a spectrum analyzer. The 
variances of the input state's  squeezed and anti-squeezed quadratures are 
$-6.2 \pm 0.2$dB and $12.0 \pm0.2$dB, respectively.  The EPR correlations 
are as good as those obtained in the teleportation of a coherent state. The 
squeezed output state is shown in Fig.~\ref{fig3}(b).  The measured 
variances of the output state are $-0.8 \pm 0.2$dB and $12.4 \pm 0.2$dB for 
the $x$ and $p$ quadratures, respectively. These values are in good
agreement with theoretical calculations, indicating a high degree of
mechanical stability in our phase-locking system. (For comparison, the
teleportation was repeated with squeezing in the $p$ quadrature, yielding
$-0.3 \pm 0.2$dB squeezing in the output.) These levels of sub-vacuum
noise measured in the output clearly demonstrate that the squeezing (and
hence the entanglement) is preserved in the process of teleportation.

Figure~\ref{fig3}(c) and (d) show the reconstructed Wigner functions of
the input and output states. Here the Wigner function is defined 
as \cite{Leonhardt_MQ(1997)}
\begin{equation}
        W(x,p) = \frac{1}{\pi}\int d\Delta \; e^{-2 i\Delta p}
      \langle x+{\Delta}/{2} |\, \hat{\rho}\, | x-{\Delta}/{2} \rangle \;.
\end{equation}
To reconstruct the Wigner function, we follow the procedures described
in Ref.~\onlinecite{Breitenbach97}. 
We record 1MHz sideband components with
a 30kHz bandwidth by an Analog-to-Digital Converter (sampling rate 300kHz,
around 100,000 points collected with the local oscillator phase
scanned). We then calculate the Wigner function by the inverse Radon
transformation \cite{Leonhardt_MQ(1997),Breitenbach97}.

In summary, we have demonstrated quantum teleportation of a squeezed state
of light that preserves the squeezing from input to output and estimate
the squeezing obtained at the output as $\Delta_{\rm sq}=0.83 \pm 0.04$.
We also show that the teleported state preserves the entanglement between 
the upper and lower sidebands. The successful preservation of entanglement 
through a teleportation channel hinges upon the existence of a 
high-fidelity teleporter, which requires high levels of squeezing 
in each of three modes. This technically challenging achievement 
opens a new door for CV quantum information processing. In particular, 
high-fidelity teleporters will enable the transport, preservation and 
manipulation of highly non-Gaussian
states \cite{Gottesman01,Furusawa07} such as single-photon states 
or superpositions of pairs of coherent states 
(sometimes called Schr\"odinger kittens)
\cite{Ourjoumtsev06,Jonas06,Wakui06}.

\vskip 0.02truein
\noindent
This work was partly supported by the MEXT of Japan. SLB currently holds
a Wolfson -- Royal Society Research Merit Award.


\begin{thebibliography}{}

       \bibitem{Mertz90}
   	J.\ Mertz {\it et al.\/}
   	{\it Phys.\ Rev.\ Lett.\/}  {\bf 64}, 2897 (1990).

        \bibitem{Josse06}
 	V.\ Josse {\it et al.\/}
 	{\it Phys.\ Rev.\ Lett.\/}  {\bf 96}, 163602 (2006).

       \bibitem{Yoshikawa07}
   	J.\ Yoshikawa {\it et al.\/}
   	{\it Phys.\ Rev.\ A }  {\bf 76}, 060301 (2007).

       \bibitem{Filip05}
   	R.\ Filip, P.\ Marek and U.L.\ Andersen,
   	{\it Phys.\ Rev.\ A }  {\bf 71}, 042308 (2005).

       \bibitem{Bennett93}
   	C.H.\ Bennett {\it et al.\/}
    	{\it Phys.\ Rev.\ Lett.\/}  {\bf 70}, 1895 (1993).

       \bibitem{Vaidman94}
        L.\ Vaidman,
    	{\it Phys.\ Rev.\ A }  {\bf 49}, 1473 (1994).

       \bibitem{Gottesman99}
   	D.\ Gottesman and I.L.\ Chuang
   	{\it Nature} {\bf 402}, 390 (1999).

    	\bibitem{Furusawa07}
    	A.\ Furusawa and N.\ Takei,
    	{\it Phys.\ Rep.\/} {\bf 443}, 97 (2007).

        \bibitem{Gottesman01}
    	D.\ Gottesman, A.\ Kitaev, J.\ Preskill,
    	{\it Phys.\ Rev.\ A} {\bf 64}, 012310 (2001).

        \bibitem{topten}
    	{\it Science} {\bf 282}, 2157 (1998).

        \bibitem{Furusawa98}
    	A.\ Furusawa {\it et al.\/}
    	{\it Science} {\bf 282}, 706 (1998).

        \bibitem{Bowen03a}
    	W.P.\ Bowen {\it et al.\/}
    	{\it Phys.\ Rev.\ A} {\bf 67}, 032302 (2003).

        \bibitem{Zhang03}
    	T.C.\ Zhang {\it et al.\/}
    	{\it Phys.\ Rev.\ A} {\bf 67}, 033802 (2003).

        \bibitem{Takei05e}
    	N.\ Takei {\it et al.\/}
    	{\it Phys.\ Rev.\ Lett.\/} {\bf 94}, 220502 (2005).

        \bibitem{Yonezawa04}
    	H.\ Yonezawa, T.\ Aoki and A.\ Furusawa,
    	{\it Nature} {\bf 431}, 430 (2004).

        \bibitem{Sherson06}
    	J.F.\ Sherson {\it et al.\/}
    	{\it Nature} {\bf 443}, 557 (2006).

        \bibitem{Takei05s}
    	N.\ Takei {\it et al.\/}
    	{\it Phys.\ Rev.\ A} {\bf 72}, 042304 (2005).

        \bibitem{JZhang03}
    	J.\ Zhang,
    	{\it Phys.\ Rev.\ A} {\bf 67}, 054302 (2003).

        \bibitem{Huntington02}
    	E.H.\ Huntington and T.C.\ Ralph,
    	{\it J.\ Opt.\ B}
    	{\bf 4}, 123 (2002).

        \bibitem{Braunstein00}
    	S.L.\ Braunstein {\it et al.\/}
    	{\it J.\ Mod.\ Opt.\/} {\bf 47}, 267 (2000).

        \bibitem{Takeno07}
    	Y.\ Takeno {\it et al.\/}
    	{\it Opt.\ Exp.\/} {\bf 15}, 4321 (2007).

        \bibitem{Duan00}
    	L.M.\ Duan {\it et al.\/}
    	{\it Phys.\ Rev.\ Lett.\/} {\bf 84}, 2722 (2000).

        \bibitem{footnoteBB}
          All broadband quantities involve an implicit integration over
        a small bandwidth --- the detector resolution --- and are
        normalized with respect to one-half of the vacuum noise level
        --- except when quoted in dB, where they are relative to the vacuum.

        \bibitem{SpectrumAnalyser}
    	R.A.\ Witte,
    	{\it Spectrum and Network Measurements}
    	(Prentice-Hall, New Jersey, 1993).

        \bibitem{vanLoock00}
    	P.\ van Loock, S.L.\ Braunstein and H.J.\ Kimble,
    	{\it Phys.\ Rev.\ A} {\bf 62}, 022309 (2000).

        \bibitem{Braunstein98}
    	S.L.\ Braunstein and H.J.\ Kimble,
    	{\it Phys.\ Rev.\ Lett.\/} {\bf 80}, 869 (1998).

       \bibitem{Suzuki06}
    	S.\ Suzuki {\it et al.\/}
    	{\it Appl.\ Phys.\ Lett.\/} {\bf 89}, 061116 (2006).

        \bibitem{Braunstein01}
    	S.L.\ Braunstein {\it et al.\/}
    	{\it Phys.\ Rev.\ A} {\bf 64}, 022321 (2001).

        \bibitem{Hammerer05}
    	K.\ Hammerer {\it et al.\/}
   	{\it Phys.\ Rev.\ Lett.\/} {\bf 94}, 150503 (2005).

        \bibitem{Grosshans01}
    	F.\ Grosshans and P.\ Grangier,
    	{\it Phys.\ Rev.\ A} {\bf 64}, 010301(R) (2001).

        \bibitem{Leonhardt_MQ(1997)}
    	U.\ Leonhardt,
    	\textit{Measuring the Quantum State of Light}
    	(Cambridge University Press, Cambridge, 1997).

        \bibitem{Breitenbach97}
    	G.\ Breitenbach and S.\ Schiller,
    	{\it J.\ Mod.\ Opt.\/} {\bf 44} 2207 (1997).

        \bibitem{Ourjoumtsev06}
    	A.\ Ourjoumtsev {\it et al.\/}
    	{\it Science} {\bf 312}, 83 (2006).

        \bibitem{Jonas06}
    	J.S.\ Neergaard-Nielsen {\it et al.\/}
    	{\it Phys.\ Rev.\ Lett.\/}  {\bf 97}, 083604 (2006).

        \bibitem{Wakui06}
    	K.\ Wakui {\it et al.\/}
    	{\it Opt.\ Exp.\/} {\bf 15} 3568 (2007).

\end{thebibliography}
\end{document}